# Tunable 2-dimensional/3-dimensional electron gases

# by submonolayer La doping of SrTiO₃

P. V. Ong,[1] Jaichan Lee[1,2] * and Warren E. Pickett[2]

[1]School of Advanced Materials Science & Engineering, Sungkyunkwan University,

Suwon, 440-746, Korea

[2]Department of Physics, University of California, Davis, California 95616, USA

ABSTRACT

First-principles calculation was used to study the structural and electronic features of the low dimensional oxide structure, $SrTiO_3/Sr_{1-x}La_xTiO_3$ (x=0.25) superlattices, constructed by submonolayer low dimensional La doping into $SrTiO_3$. We demonstrate a dimensionality crossover from three-dimensional to two-dimensional (3D → 2D) electronic behavior in the system. Two types of carriers, one confined to 2D and the other extended, exhibit distinct tunable (3D → 2D) transport characteristics that will enable the study of many properties (e.g., superconductivity) through this change in dimensionality.

Transition metal oxides with partially filled $d$ orbitals, when under spatial confinement, display unusual properties arising from the interplay of strong correlation and quantum confinement effects. When hybridization with oxygen $2p$ orbitals is also varied, various internal degrees of freedom combine or compete in the formation of collective order.[1] Another external parameter, the carrier doping level, plays a central role,[2,3] leading to superconductivity, spin order, and orbital order in oxide nanostructures. Yet another variable to tune is the dimensionality together with the concentration of dopants: confining dopants to layers rather than being randomly distributed, leading to anisotropy of properties as well as potentially to enhanced fluctuations of the several degrees of freedom. Recent researches on perovskite oxide heterostructures have mainly focused on LaAlO$_3$/SrTiO$_3$ (LAO/STO)[4-7] or LaTiO$_3$/SrTiO$_3$ (LTO/STO)[8] systems, that exhibited several intriguing properties associated two dimensional electron gas (2DEG), i.e., high mobility,[4] magnetism,[5] or superconductivity.[6,7]

In this paper we propose an artificially grown structure to change the dimensionality of a perovskite oxide through submonolayer doping of single layers into the perovskite oxide. The low dimensional structure produces a heretofore unseen 3DEG to 2DEG crossover that may provide novel functionalities for oxide nanostructures. The most well known generic prototype of such a system is high $T_c$ superconducting cuprates, in which charge carriers are doped into the Cu-O plane from insulating block layers. On the other hand, the low dimensional chemical doping proposed here realizes a tunable dimensionality accompanying the changing band filling. Our earlier study reported successful growth of low dimensional (2D) chemically doped SrTiO$_3$ oxide, which could be repeated to produce superlattices,[9] for which electron transport behavior differed greatly from bulk.[10] A similar low dimensional oxide superlattice (Nb doping rather than La) was reported to exhibit a giant Seebeck coefficient, interpreted as arising from the 2D electron gas.[11] Such an embedded 2D electron gas was shown to exhibit superconductivity with quantum oscillations of superimposed frequencies in the normal state.[12] On the other hand, no theoretical studies of submonolayer-doped oxide nanostructures exist, perhaps because complete-layer oxide heterostructures (e.g., LTO/STO) have been drawing the most attention.[13,14]

In this study, we use first-principles calculations on $(SrTiO_3)_n/(Sr_{1-x}La_xTiO_3)_1$ superlattices ($n \times 1$ STO/SLTO SL with x=0.25) and demonstrate that (1) the three-fold degenerate Ti $t_{2g}$ orbitals split into $d_{xy}$ and ($d_{xz}, d_{yz}$), whose spacing in energy is tunable by the STO spacer thickness ($n$), (2) carriers occupying the multiple subbands possess distinctive transport characteristics, i.e., some have large effective mass and are strongly localized at the interfaces, the others show high mobility and well extended in the whole system, and (3) change in the repeat distance ($n+1$) (which alters the dimensionality) tunes both the transport characteristics and the relative concentrations of the 2D vs. 3D carrier density, inducing a dimensional crossover. This behavior contrasts sharply with bulk doping with its isotropic conduction. Our results also provide an explanation of recent experimental results for the 2DEG in STO, i.e., multichannel conduction of LTO/STO SL in a delta-doping geometry.[15]

The $n \times 1$ SL is studied using a $2 \times 2 \times (n+1)$ supercell which is composed of a single monolayer of $Sr_{0.75}La_{0.25}TiO_3$ embedded in STO every $n$ layers, just as has been accomplished with PLD.[9] The in-plane lattice parameter is fixed at $2a_{STO}$ ($a_{STO}$ = 3.895Å is the calculated lattice constant of cubic STO) to simulate the mechanical boundary condition imposed by the STO substrate. First-principles density functional calculations were performed within the projector augmented wave (PAW) method[16] as implemented in the VASP code[17,18] with $4\times4\times2$ $k$-point grid and an energy cutoff of 500eV. For comparison, bulk SLTO was also modeled by a $3\times3\times3$ supercell with one single La substituting for a Sr atom (3.7%. doping). To treat exchange and correlation effects, we use the LDA+$U$ method with $U_d$ = 5 eV and $J_d$ = 0.64 eV for the Ti $d$ states and $U_f$ = 11 eV and $J_f$ = 0.68 eV for La $f$ states. The large value of $U_f$ for La was used to shift the La $f$ band further up in energy and prevent a spurious mixing with the Ti $d$ bands.[19,20] A full structural optimization was performed within the tetragonal symmetry of the STO cell, i.e. the atomic displacements are allowed exclusively along the c-axis. However, further structural distortion, i.e., antiferrodistortion of STO, was also investigated to examine the effect of antiferrodistortion on band structure and transport. Lattice relaxation processes were performed until forces acting on atoms were less than 0.01eV/Å.

Substitution of La for Sr provides one extra electron that is transferred to surrounding Ti 3$d$ states. Upon the low dimensional La substitution, 2D arrangement of the trivalent La dopants forms a positively charged La-sheet, which must be screened, both by electrons and by ions. For the electronic screening, the positive La layer produces a potential well, near which the electrons will be confined if the layer is isolated. Ionic screening occurs through atomic relaxation: Ti and Sr cations move outward from the positively charged La-sheet while O anions move inward. Such a polar distortion produces local dipoles pointing outward from the La layer, displayed for the $5\times1$ SL in Fig. 1a, which decreases monatonically away from the La layer. Similar polar distortions have been reported for LTO/STO and LAO/STO heterostructures.[13,14,19,21] The octahedral distortion can be quantified by the local octahedron dimension $d = \left[ z(\mathrm{O_{Ib}}) - z(\mathrm{O_{Ia}}) \right]$ along the *z*-direction. The variation of *d* with the octahedron position is shown in Fig. 1b for all systems we have studied. For each $n \times 1$ stacking sequence, the apical length *d* of the octahedron bordering the La doped layer is shorter than for subsequent layers. With increasing doping layer separation (thicker STO slabs), the apical length *d* of the octahedron adjacent to the La doped layer initially decreases toward that of an undistorted octahedron (*d*=3.895Å, calculated) but then crosses and becomes smaller than the undistorted *n*>3 unit cells. The observation that this octahedron becomes compressed for *n*>3 n correlates with experimental results on the same system,[10] where precisely this value is the critical thickness at which a metal-insulator transition occurs. Thus the compression of the octahedron appears to be associated with the transition, as well as the changing (average) carrier density. An analogous compression near the interface was also experimentally reported in the LAO/STO system.[22]

Now we describe the tuning of the electronic system afforded by the layer variable *n*. Figure 2 shows the near-Fermi energy bands and projected density of states (PDOS) of the Ti 3*d*-derived $t_{2g}$ ($d_{xy}$, $d_{yz}$ $d_{xz}$) for bulk STO and for *n* = 1 and 5. The PDOS is plotted for the Ti site adjacent to the La-doped plane. The threefold degeneracy of the $t_{2g}$ manifold is removed due to the presence of the La doped layer and to the structural distortion discussed above (Fig. 1). In the *n*=1 case, the $d_{xy}$ orbital is hardly split off (by no more than 10meV, Fig. 2), but for *n*=5 the orbital reconstruction is clear (40meV splitting). This magnitude of splitting is sufficient that the $d_{yz}$, $d_{xz}$ orbitals are emptied: the transition to a pure $d_{xy}$ band occupancy. Similar 3*d* band splitting has been observed in the LAO/STO heterostructures, where x-ray absorption spectroscopy showed that the $d_{xy}$ orbital has a lower energy than ($d_{yz}$, $d_{xz}$) by 50meV and becomes the first available states for conducting electrons.[23] Recently, Kozuka *et al.*[12] have shown that Shubnikov-de Haas oscillations of superimposed frequencies were observed in a similar system, i.e., Nb-doped STO embedded in STO, which could be attributed to a confined electron gas of multiple subbands.

To characterize the transport properties of the conducting electrons, we consider the dispersion of the multiple sub-bands in more detail. As is seen in Fig. 2, the $d_{xz}$, $d_{yz}$ bands still show a parabolic dispersion near the $\Gamma$ point while the $d_{xy}$ band becomes narrow along the out-of-plane $\Gamma$-Z direction for $n$=1. Correspondingly, the width of the $d_{xy}$ PDOS peak is significantly decreased in comparison with that in bulk. For $n$=5 the lowest $d_{xy}$ band becomes very flat along the $\Gamma$-Z direction and the width of its PDOS peak is further reduced. Independently of $n$, the band in-plane dispersion along $\Gamma$-X direction is almost unchanged. The strongly reduced band dispersion along the $\Gamma$-Z direction and width of the PDOS peak result in increasing electron confinement of only one of the multiple types of carriers, and the preferential occupation of the $d_{xy}$ orbital provides an optimized electronic structure for an enhanced Seebeck coefficient $S$.[11,24] Mahan-Sofo theory[24] suggests that the best thermoelectric material is likely to be found among materials exhibiting a local increase in the DOS over a narrow energy range. We also examined the effect of the antiferrodistortive mode, in which neighboring $TiO_6$ octahedra rotate oppositely about the c-axis.[25] The result obtained for a 3x1 SL shows the same spitting pattern and band dispersion as those in the present case, i.e. the singly degenerate $d_{xy}$ is significantly lower than the doubly degenerate ($d_{yz}$,$d_{xy}$), weak dispersion along the $\Gamma$-Z direction and predominantly occupied by carriers. These electronic behavior has recently been reported for STO in antiferrodistortive phase.[26]

We have calculated within the parabolic approximation the carrier effective masses near the $\Gamma$ point, with results shown in Table I. The effective masses of the ($d_{yz}$, $d_{xz}$) electrons and the in-plane effective mass of the $d_{xy}$ electrons hardly change with stacking period. However, the out-of-plane effective mass of the $d_{xy}$ carriers increases strongly with $n$, consistent with the band dispersion (Fig. 2). DOS effective mass $m_d$ was also calculated since $S$ is expressed in terms of DOS effective mass for a degenerate semiconductor.[27] For $n=5$, where the $d_{xy}$ orbital is exclusively occupied, $m_d =6.7m_0$ was obtained for the $d_{xy}$ electrons and can be compared with the estimated value $m_d =9.3m_0$,[28] for a $(SrTi_{0.8}Nb_{0.2}O_3)/(STO)_9$ SL that was reported to have giant thermoelectric (Seebeck) coefficient. In addition, the distinctive lattice distortion and presence of the La dopant near the interface could further reduce thermal conductivity by phonon scattering, and hence enhance thermoelectric figure of merit.[29] Therefore, both electronic reconstruction and the structural distortion found in this STO/SLTO SL would contribute to an enhanced thermoelectric functionality.

The band structure and effective mass calculation indicate two distinct types of conducting electrons in this electronic system embedded in STO, with each type being distinguished by its transport characteristics. Figure 3 shows the band-decomposed density contours of the electrons at the bottom of the conduction band in a $5\times1$ SL before and after ionic relaxation. For this large stacking period, most of the carriers reside in the $d_{xy}$ orbital, consistent with the electronic structure results (Fig. 2). Before the relaxation, the electrons in all the split-off bands are mostly localized on the Ti layers nearest to the SLO layer (Fig. 3, *left*). After the relaxation, the screening associated with the ionic polar distortion reduces the depth of the potential-well caused by $La^{+3}$ in the SLO layer. The local variation of dipole moment is screened by the doped electrons, resulting in the further spreading of the electrons into the STO spacer (Fig. 3, *right*). Despite the screening by the lattice, the $d_{xy}$ electrons spread very little, remaining primarily on the Ti ions neighboring the La doped layer. On the other hand, the carriers with significantly lower density in the doubly degenerate $d_{xz}$ and $d_{yz}$ orbitals spread to become homogeneously distributed over the whole system. As a result, as dimensionality is lowered, the $d_{xy}$ electrons would comprise the main fraction of carriers in the embedded Q2-DEG. Such tunable 2D electron confinement can induce the early onset of electron correlation compared with bulk doping and will be critical in determining further transport properties associated with the Q2-DEG. At the completion of this work, we became aware of new experimental results on 2D transport properties of STO/LTO SL. Kim *et al.*[15] showed that the transport behavior of the STO/LTO SL is well explained by a "two-carrier model": one possesses high density and low mobility, the other low density and high mobility. This framework model is completely consistent with our predictions.

In summary, we have shown that two dimensional submonolayer electron doping of $SrTiO_3$ produces a hybrid density of $d_{xy}$ 2D carriers is embedded in a much lower density 3D conductor. The relative carrier densities can be tuned by lowering dimensionality (i.e., increasing the separation of the doping layers), effectively spanning the complete range from 3D to 2D conduction through a variable anisotropy regime. These results provide an explanation and understanding of recent observations, i.e. high thermoelectric performance in Nb:STO/STO SL and the "two-carrier model" for the 2D transport behavior STO/LTO system, and may inspire further investigations on tunable low dimensional oxide structures.

This work is supported by the Basic Science Research Program through National Research Foundation of Korea (2009-0092809).  W.E.P. was supported by DOE Grant DE-FG02-04ER46111.

Figure captions

Figure. 1. (Color online) (a) Optimized atomic structure for the $5 \times 1$ SL. The signed numbers (in Å) indicate the displacement from the unrelaxed distances along z-axis between Ti and equatorial $O_{II}$'s (pink), Ti and the apical $O_{Ia}$'s or $O_{Ib}$'s (blue), and apical $O_{Ia}$'s and $O_{Ib}$'s (red). The positive (negative) sign means increase (decrease). (b) Distortion of the $TiO_6$ octahedra on both sides of the SLO layer in the $n \times 1$ SL ($n$=1-5) (a). The vertical dashed line indicates the undistorted octahedron.

Figure 2. (Color online) The band structure at the bottom of the conduction band (left) and PDOS (right) of the Ti $3d$-derived $t_{2g}$ ($d_{xy}$, $d_{yz}$, and $d_{xz}$) orbitals for bulk $Sr_{1-x}La_xTiO_3$, $1 \times 1$(n=1), and $5 \times 1$ (n=5) SL. The PDOS is plotted for the Ti atoms nearest to the La-doped (SLO) plane in the SL.

Figure 3. (Color online) Band-decomposed charge densities of the conducting electrons for a $5 \times 1$ superlattice. The contours are plotted in a (010) plane passing through Ti ions before relaxation (left) and after relaxation (right). The unit of contour values is e/Å$^3$.

TABLE I: Calculated effective masses of electrons in $d_{xy}$ and ($d_{xy}$, $d_{xz}$) orbitals in $(STO)_n/(SLTO)_1$ superlattices ($n$=1-5). The symbols of $m_\parallel$, $m_\perp$ and $m_d$ denote effective mass along $\Gamma$-X, $\Gamma$-Z and density-of-states effective mass, respectively. All values are in units of a free-electron mass $m_0$.

| Stacking period | $d_{yz}$ or $d_{xz}$ | | | $d_{xy}$ | | |
|---|---|---|---|---|---|---|
| | $m_\parallel$ | $m_\perp$ | $m_d$[a] | $m_\parallel$ | $m_\perp$ | $m_d$[a] |
| 1x1 | 0.4 | 0.5 | 0.4 | 4.0 | 4.3 | 4.1 |
| 2x1 | 0.4 | 0.5 | 0.4 | 3.8 | 5.7 | 4.3 |
| 3x1 | 0.4 | 0.5 | 0.4 | 4.0 | 8.6 | 5.2 |
| 4x1 | 0.4 | 0.5 | 0.4 | 3.9 | 13.7 | 5.9 |
| 5x1 | 0.4 | 0.5 | 0.4 | 3.9 | 20.1 | 6.7 |

[a] Density-of-states effective mass is estimated according to $m_d = (m_\perp m_\parallel^2)^{1/3}$

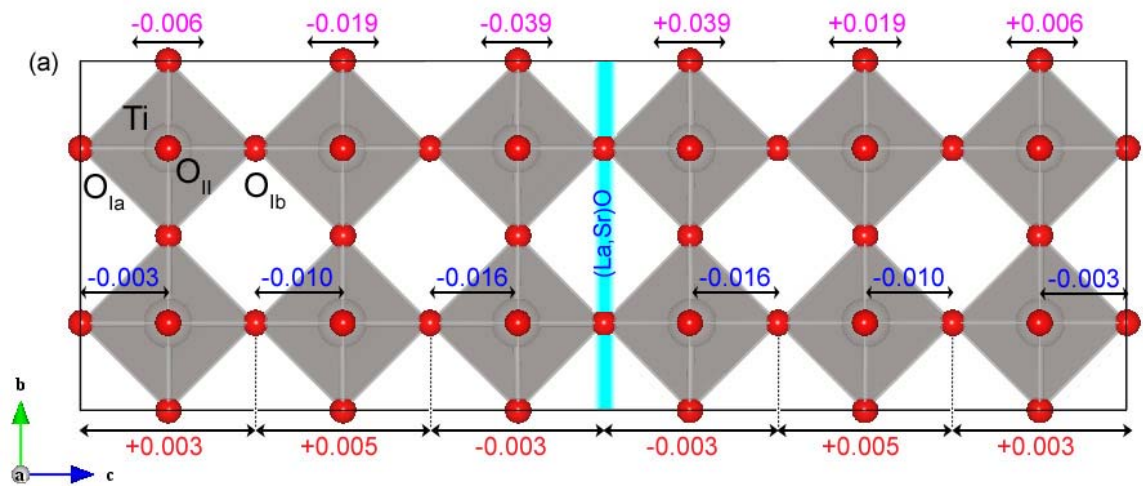

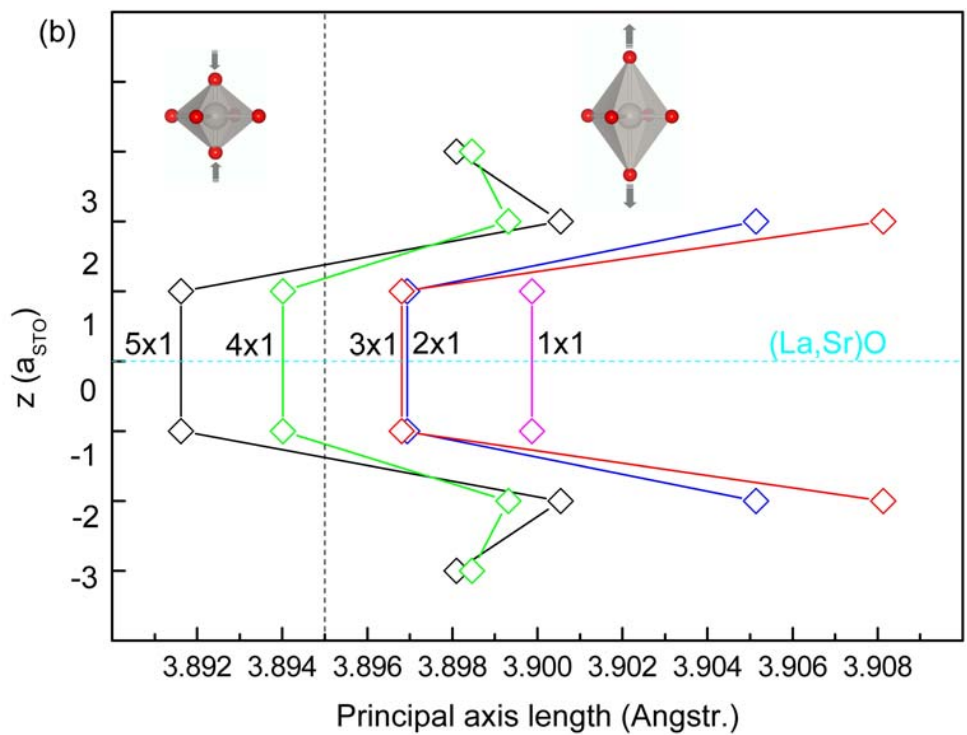

Figure 1.

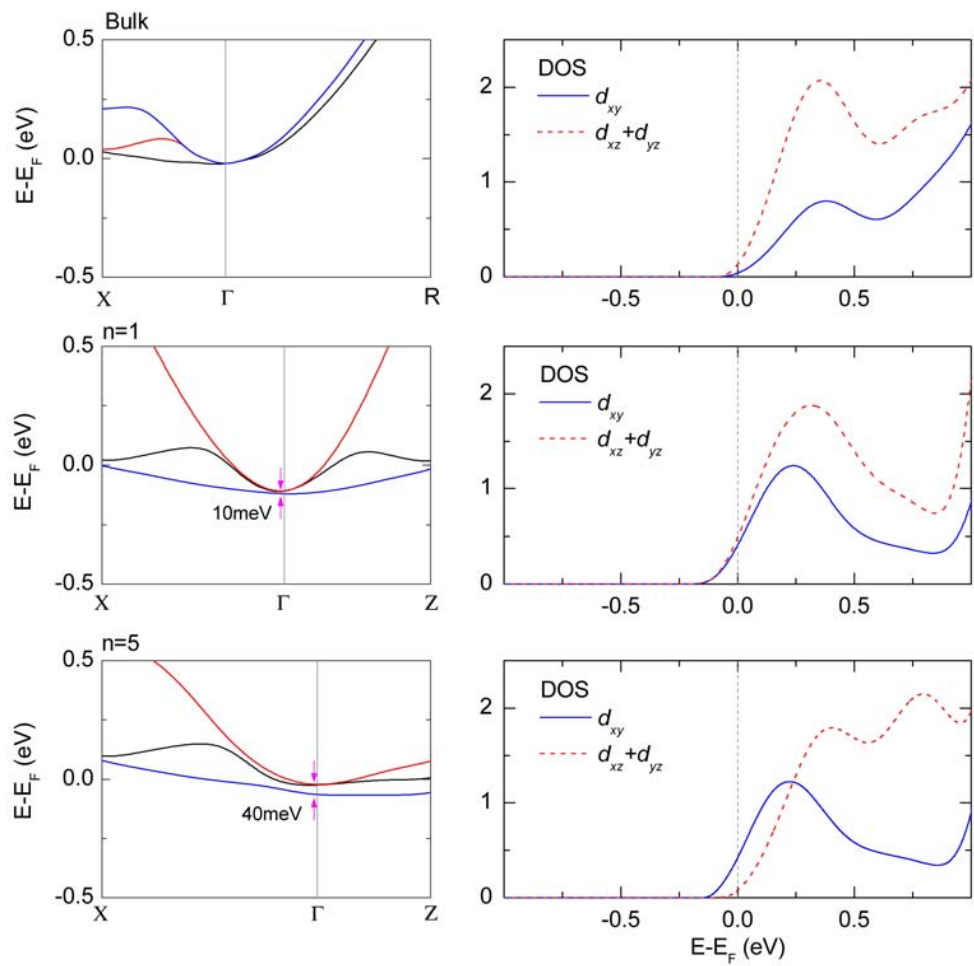

Figure 2

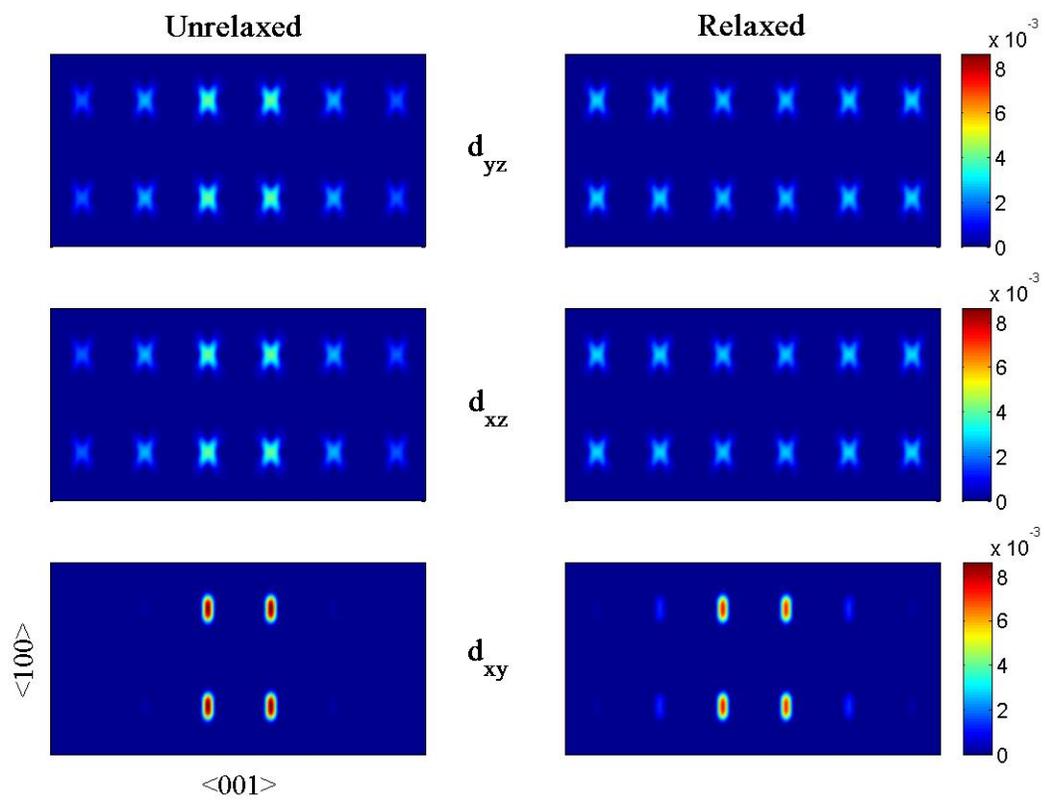

Figure 3